\documentclass[aps,amsmath,amssymb,twocolumn]{revtex4}

\usepackage{graphicx}
\usepackage{dcolumn}
\usepackage{bm}
\usepackage{amsmath}
\usepackage{epstopdf}
\usepackage{amsfonts}
\usepackage{amssymb}
\usepackage{epsfig}
\usepackage{tabularx}
\usepackage{color}
\usepackage{soul}
\usepackage[dvipsnames]{xcolor} 

\usepackage[colorlinks=true,citecolor=blue,urlcolor=Magenta,breaklinks]{hyperref}

\newcommand{\be}{\begin{equation}}
\newcommand{\en}{\end{equation}}
\newcommand{\bea}{\begin{eqnarray}}
\newcommand{\ena}{\end{eqnarray}}

\newcommand{\Wr}{ \Omega_r }
\newcommand{\Wm}{ \Omega_m }
\newcommand{\Wl}{ \Omega_\Lambda }

\begin{document}

\title{Growth of structures and redshift-space distortion data in scale-dependent gravity}

\author{
Grigoris Panotopoulos  {${}^{a}$
\footnote{\href{mailto:grigorios.panotopoulos@tecnico.ulisboa.pt}{grigorios.panotopoulos@tecnico.ulisboa.pt} }
}
\'Angel Rinc\'on {${}^{b}$
\footnote{
\href{mailto:aerinconr@academicos.uta.cl}{aerinconr@academicos.uta.cl}}
} 
}

\address{
${}^a$ Centro de Astrof{\'i}sica e Gravita{\c c}{\~a}o-CENTRA, Instituto Superior T{\'e}cnico-IST, Universidade de Lisboa-UL, Av. Rovisco Pais, 1049-001 Lisboa, Portugal.   
\\
${}^b$ Sede Esmeralda, Universidad de Tarapac\'a,
Avda. Luis Emilio Recabarren 2477, Iquique, Chile.
}

\begin{abstract}
This study is devoted to the implications of scale-dependent gravity in Cosmology. Redshift-space distortion data indicate that there is a tension between $\Lambda$CDM and available observations as far as the value of the rms density fluctuation, $\sigma_8$, is concerned. It has been pointed out that this tension may be alleviated in alternative theories in which gravity is weaker at red-shift $z \sim 1$. We study the evolution of density perturbations for non-relativistic matter on top of a spatially flat FLRW Universe, and we compute the combination $A=f \sigma_8$ in the framework of scale-dependent gravity, where both Newton's constant and the cosmological constant are allowed to vary with time. 
Upon comparison between available observational data (supernovae data as well as redshift-space distortion data)
and theoretical predictions of the model, we determine the numerical value of $\sigma_8$ that best fits the data.
\end{abstract}

\maketitle

\section{Introduction}

Current data seem to suggest that we live in a spatially flat Universe dominated by dark matter and dark energy \cite{turner}. Although as of today the origin and nature of the dark sector still remains unknown, we have come up with the standard cosmological model, which is based on cold dark matter combined with a positive cosmological constant ($\Lambda$CDM) within Einstein's General Relativity (GR) \cite{GR}. This model, despite its simplicity, is overall in a very good agreement with a wealth of observational data coming from Astrophysics and Cosmology. As economic and successful as it may be, the concordance cosmological model unfortunately does not come without problems. 

\smallskip

To begin with, as far as the theory itself is concerned, i) it contains singularities that indicate that the theory is incomplete, and ii) it is a non-renormalizable theory. In addition to those, $\Lambda$CDM, too, comes with its own problems. In particular, apart from the well-known fine-tuning problem, there are a couple of tensions related to $\Lambda$CDM discussed in the literature over the last years. To be more precise, there is nowadays a tension regarding the value of the Hubble constant, $H_0$, between high red-shift CMB data and low red-shift data, see e.g. \cite{tension,tension1,tension2,tension3}. The value of the Hubble constant 
extracted by the Planck Collaboration \cite{planck1,planck2}, $H_0 = (67-68)~\text{km/(Mpc  sec)}$, is found to be lower than the value obtained by local measurements, $H_0 = (73-74)~\text{km/(Mpc sec)}$ \cite{hubble,recent}. What is more, regarding large scale structure formation data, the growth rate from red-shift space distortion measurements has been found to be lower than expected from Planck \cite{eriksen,basilakos}.

\smallskip

Regarding the aforementioned problems and possible alternatives to the $\Lambda$CDM model, either a modified theory of gravity is assumed, providing correction terms to GR at cosmological scales, or a new dynamical degree of freedom with an equation-of-state (EOS) parameter $w < -1/3$ must be introduced. In the first class of models (geometrical DE) one finds for instance $f(R)$ theories of gravity \cite{mod1,mod2,HS,starobinsky} (see also \cite{NewRef1,NewRef2} for a more general class of theories, $f(R,L)$, that allow for more general couplings between matter and curvature), brane-world models \cite{langlois,maartens,dgp} and Scalar-Tensor theories of gravity \cite{BD1,BD2,leandros1}, while in the second class (dynamical DE) one finds models such as quintessence \cite{DE1}, phantom \cite{DE2}, quintom \cite{DE3}, tachyonic \cite{DE4} or k-essence \cite{DE5}. What is more, people have proposed and analysed the implications of cosmological models with a varying cosmological constant, see e.g. \cite{Sola1,Sola2,Sola3,Sola4} and references therein. The effective cosmological equations in $\Lambda$-varying scenarios can be computed without additional assumptions, they are a natural generalization of the more standard concordance model, and they offer a richer phenomenology compared to the 
$\Lambda$CDM model \cite{Oztas:2018jsu}. 
There are also some works on cosmological models with a variable Newton's constant, see e.g. \cite{VarG1,VarG2,VarG3}.

\smallskip

Over the last years scale-dependent gravity has emerged as an interesting framework with a few appealing properties. As it is inspired by the renormalization group approach, it naturally allows for a varying cosmological constant and a varying Newton's constant at the same time. Its impact on black hole physics has been investigated in detail \cite{SD1,SD0,SD2,SD3,SD4,SD5,SD6,SD7,SD8}, while recently some astrophysical and cosmological implications have been studied as well \cite{astro,cosmo1,cosmo2}. 
Recently, in \cite{cosmo2} the authors considered current cosmic acceleration within scale-dependent gravity,
analyzing in particular the evolution of the Universe at the background level. In the present work we propose to study for the first time the evolution of linear matter perturbations, and in particular the predictions for growth rate from red-shift space distortions. Since scale-dependent gravity applied to cosmological contexts implies a varying Newton´s constant, $G_{\textrm{eff}}$, which is lower than ordinary Newton's constant, $G$, at red-shift $z \sim (1-2)$, it would be both natural and interesting to investigate the predictions of scale dependent gravity on the $A$ parameter, see the discussion below in section 4.

\smallskip

The paper is organized as follows: In the next section we briefly describe the formalism in the framework of scale-dependent gravity. It is then applied to Cosmology in section 3, where we present the cosmological equations for classical and scale-dependent background evolution. In the fourth section, where linear cosmological perturbation theory is discussed, we study the evolution of density perturbations for pressure-less matter, and we compute the combination $A=f \sigma_8$. Finally, we finish with some concluding remarks in section 5.

\section{Scale-dependent gravity}

The goal of this section is to briefly review the key concepts and ingredients of scale-dependent gravity (see e.g. \cite{SD0,SD8} and references therein). Scale-dependent (SD) gravity is one of the current alternative scenarios in which the couplings of the underlying theory are allowed to vary. The motivation for studying alternatives scenarios is, roughly speaking, to address a few shortcomings of the $\Lambda$CDM model. In particular, the $H_0$ tension as well as the cosmological constant problem maybe alleviated when the couplings of the theory vary with time. Among other formalisms, scale-dependent gravity is one where a soft entanglement between classical and quantum is established. To be more precise, the technical motivation of this approach relies on the idea of quantum gravity \cite{Rovelli:2007uwt}. Exact Renormalization Groups (ERG) is one of the now many alternatives to quantum gravity, and it serves as an inspiration. Currently, we do not have a favorite approach to combine classical gravity to quantum mechanics and, therefore, a "consistent and predictive" parameterization of quantum gravity is still missing. Scale-dependent gravity takes advantage of certain ideas of ERG to consistently account for quantum effects. Thus, considering an average effective action with running couplings, a successful theory can be achieved. Also, it is remarkable that the ERG and the SD scenario allow us to obtain the equations for running couplings of an average effective action exactly. Thus, the latter approaches do not need additional expansion of the couplings in powers of some small control parameter. 

\smallskip

Scale-dependent gravity is able to extend classical GR solutions via the inclusion of scale-dependent couplings, which account for quantum features. Notice that the corrections are assumed to be small. The idea can also be shown considering the truncated average effective Einstein-Hilbert action with a non-vanishing cosmological constant. Thus, in the absence of matter content, there are two "running" couplings in the theory, namely: i) Newton's constant $G_k$ and ii) the cosmological constant $\Lambda_k$. For convenience, we can define the parameter $\kappa_k \equiv 8 \pi G_{k}$ as the Einstein's constant. 
Besides, apart from the the metric tensor, $g_{\mu \nu}$, an arbitrary renormalization scale $k$ must be included.

\smallskip

The effective action $\Gamma[g_{\mu \nu}, k]$ is then written as \cite{SD0,SD8}
\begin{equation}
	\Gamma[g_{\mu \nu}, k] \equiv \int \mathrm{d}^4 x \sqrt{-g}
	\Bigg[ 
	\frac{1}{2 \kappa_k} \Bigl(\mathcal{R} - 2 \Lambda_k \Bigl) \ + \ \mathcal{L}_M
	\Bigg],
\end{equation}
where $\mathcal{L}_M$ is the Lagrangian density of the matter fields, $g$ is the determinant of the metric tensor $g_{\mu \nu}$,  $\Lambda_k$ is the scale-dependent cosmological constant (CC), $\mathcal{R}$ is the corresponding Ricci scalar and $\kappa_k$ is the scale-dependent gravitational coupling. To obtain the field equations we vary (the average effective action) with respect to the metric tensor. The latter gives the effective Einstein's field equations, which can be written down as:
\begin{equation}
G_{\mu \nu } + \Lambda_k g_{\mu \nu} \equiv \kappa_k T_{\mu \nu}^{\text{effec}},
\end{equation}
where the effective energy-momentum tensor is computed to be
\begin{equation}
\kappa_k T_{\mu \nu}^{\text{effec}} =  \kappa_k T_{\mu \nu}^{M} - \Delta t_{\mu \nu}.
\end{equation}
Notice that the  effective energy-momentum tensor includes two parts: i) the usual matter content and ii) the non-matter source (provided by the running of the gravitational coupling), which is given by \cite{SD9}:
\begin{equation}
\Delta t_{\mu \nu} \equiv G_k \Bigl( g_{\mu \nu} \square - \nabla_{\mu} \nabla_{\nu} \Bigl) G_k^{-1}. 
\end{equation}
In the present work we propose to investigate the impact of scale-dependent gravity on linear cosmological perturbations including a non-vanishing cosmological constant.

\smallskip

The second step to obtain the full set of equations consists of taking the variation of the average effective action with respect to the additional field $k(x)$. Doing so, we obtain a supplementary equation to close the set of equations. Thus, such a condition can be written down as 
\begin{equation}\label{eomk}
\frac{\delta \Gamma[g_{\mu \nu},k]}{\delta k} \bigg|_{k=k_{\text{opt}}} = 0.
\end{equation}
The latter is taken as an aposteriori condition towards background independence \cite{Stevenson:1981vj,Reuter:2003ca,Becker:2014qya,Dietz:2015owa,Labus:2016lkh,Morris:2016spn,Ohta:2017dsq}. Also, using \eqref{eomk} we are able to link  $G_k$ 
with $\Lambda_k$. The system is now closed after the inclusion of the aforementioned equation.
The auxiliary scalar field $k(x)$ is a real physical scale, identified with the momentum as a function of any space-time point $x^\mu$. The way in which the theory depends on $k$ is accounted for via the running of the coupling constants.   
The dependence of $k$ on the physical coordinates is not unique. 
Thus, an alternative choice to circumvent such a problem could be the following: First we recognize that 
$
\mathcal{O}(k(x)) \rightarrow \mathcal{O}(x)
$,
and then we can forget about the concrete relation between $k$ and the spatial coordinates.

Finally, it is instructive to include one additional equation \cite{Koch:2010nn}
\begin{align}
- \Bigl( \mathcal{R} - 2 \Lambda \Bigl) \frac{\mathrm{d}}{\mathrm{d}k}\ln(G) = 2 \frac{\mathrm{d}\Lambda}{\mathrm{d}k}.
\end{align}
which although is not used in practice, it serves as a consistency relation that shows that a running cosmological constant implies a running Newton's coupling and vice-versa, a fact that has been previously reported in \cite{cosmo1,Cai:2011kd}.

\section{Flat FLRW Universe}

In this section we briefly present the cosmological equations for the background, first the
classical one and then its scale-dependent counterpart.

\subsection{Classical background}

This subsection is dedicated to reviewing the main features of the ordinary cosmological model within GR, see e.g. the review \cite{review1}. Such a model is based on two main assumptions: i) a homogeneous and ii) a isotropic Universe, which is described by the following line element
\begin{align}\label{lineEl}
ds^2 &= -dt^2 + a(t)^2 \delta_{ij} dx^i dx^j \,,
\end{align}
setting the curvature of the 3-space to zero, $k=0$, where $t$ is the cosmic time, and $a(t)$ is the scale factor. 
Taking the latter into account, Einstein's field equations including a non-vanishing cosmological constant are the following
\begin{equation}
G_{\mu\nu} + \Lambda g_{\mu\nu}=\kappa T_{\mu\nu}\,,
\end{equation}
where $\kappa = 8 \pi G$, with $G$ being Newton's constant, and $\Lambda$ is the cosmological constant. Also, notice that the matter content of the Universe is parameterized as a perfect fluid, with an energy-momentum tensor given by
\begin{align}
T_{\nu}^{\mu} =\text{diag}(-\rho,p,p,p)
\end{align}
where $p$ is the pressure, and $\rho$ the energy density of the cosmological fluid.
 
The cosmological equations are found to be the continuity equation as well as the two Friedmann
equations
\begin{eqnarray}
H^2 & = & \frac{8 \pi G}{3} \rho \\
\frac{\ddot{a}}{a} & = & -\frac{4 \pi G}{3} (\rho + 3p) \\
0 & = & \dot{\rho} + 3 H (\rho+p)
\end{eqnarray}
where an over dot denotes differentiation with respect to cosmic time, and $H=\dot{a}/a$ is the Hubble parameter. If there are several fluid components, then
\begin{eqnarray}
p & = & \sum_i p_i \\
\rho & = & \sum_i \rho_i
\end{eqnarray}
For barotropic fluids $p=w \rho$, where $w$ is the equation-of-state parameter. If $w$ is a constant, the continuity equation can be immediately integrated to obtain the energy density in terms of the scale factor as follows
\begin{equation}
\rho = \frac{\rho_0}{a^{3(1+w)}}
\end{equation}
where $\rho_0$ is the present value of the energy density, and $a_0=1$. For convenience, we introduce at this point the normalized densities
\begin{align}
\Omega_X &\equiv \frac{\rho_{X}}{\rho_{c}}\,, \quad \rho_c=\frac{3H_0^2}{8 \pi G}\,,
\end{align}
where $X = \{\Lambda, r, m \}$, for each fluid component. Here we consider three fluid components, namely
non-relativistic matter ($w=0$), radiation ($w=1/3$), and cosmological constant ($w=-1$). Therefore, the Hubble parameter
takes the form
\begin{equation}
H^2 = H_0^2 \left( \Wl + \frac{\Wr}{a^{4}} + \frac{\Wm}{a^{3}} \right)
\end{equation}
Introducing the dimensionless Hubble parameter, $E = H/H_0$, these equations reduce to
\begin{align}\label{Feq}
E(a)^2  &= \Wl + \frac{\Wr}{a^{4}} + \frac{\Wm}{a^{3}}\,, 
\end{align}
where all $\Omega_i$, $i=m,r,\Lambda$, correspond to the present values, dropping for simplicity the sub-index "0".
Clearly, they must satisfy the constraint 
\begin{equation}
\sum_i \Omega_i = 1
\end{equation}
which may be used to express one in terms of the others, for instance 
\begin{equation}\label{dps3}
 \Wl = 1-\Wm-\Wr\,.
\end{equation}

\subsection{Scale-dependent background}

Following \cite{cosmo2}, the corresponding effective Friedmann equations in scale-dependent gravity and the saturated version of the null energy condition acquire the following form
\begin{align}
&\frac{1}{H_0^2}\left(H^2 - H \frac{\dot g}{g}\right)=\Wl \lambda(t) + \frac{\Wr}{a^{4}} g + \frac{\Wm}{a^{3}} g \,, \label{SD1} \\
&\frac{1}{H_0^2}\left(2  \dot H + 3H^2-2H\frac{\dot g}{g}+2\frac{\dot g^2}{g^2} - \frac{\ddot g}{g} \right)=3 \Wl \lambda(t) - \frac{\Wr}{a^{4}} g\,, \label{SD2} \\ 
& \frac{\ddot g}{g} -H\frac{\dot{g}}{g} - 2 \frac{\dot g^2}{g^2}=0\,, \label{NECCosm}
\end{align}
where
\begin{equation}\label{dimless}
g(t) \equiv \frac{G(t)}{G_0}\,, \quad \lambda(t) \equiv \frac{\Lambda(t)}{\Lambda_0}\,.
\end{equation}
Notice that in the last equality we introduced dimensionless quantities, $g(t)$ for the varying Newton's constant, and 
$\lambda(t)$ for the varying cosmological constant. 

We remark in passing that in the simplified case where $\Omega_r$ and $\Omega_m$ are set to zero 
the system of coupled cosmological equations admits an exact analytical solution obtained in \cite{cosmo1}. 
In that, both Newton's coupling and the cosmological constant evolve with time. The solution is characterized by a single parameter that measures the deviation from the classical solution. When that parameter is precisely zero, both the cosmological and Newton's constant are constants and equal to their classical values. This a concrete and simple example that very nicely demonstrates that a running $G(t)$ necessarily implies a running $\Lambda(t)$.

Notice that apart from the two Friedmann equations, there is an additional differential equation for the time evolution of $G(t)=G_0 g(t)$. Since equation \eqref{NECCosm} is a second order differential equation for $g(t)$, we need to specify two initial conditions, both for $g$ itself and its first derivative, $\dot{g}$. Moreover, in the generalized cosmological equations there are some new features not present in conventional Friedmann equations, and those are the following:

\begin{enumerate}
\item On the right hand side: the cosmological constant term is there, but now it depends on time.
\item On the right hand side: in the matter and radiation terms, although they scale with the scale factor as 
usual ($1/a^3$ and $1/a^4$, respectively), there is a prefactor $g(t)$, which depends on time, and which is different than unity.
\item On the left hand side: There is a new term, $\dot{g}/g$, which expresses the impact of a varying Newton's constant on the expansion of the Universe.
\end{enumerate}
Therefore the acceleration of the Universe takes place due to a running cosmological constant as well as a running Newton's constant, or in other words the varying cosmological constant combined with the g variation play the role 
of dark energy.

\smallskip

In total there are three equations, and three unknown quantities to be determined, namely $a(t),g(t),\lambda(t)$. Note that $\lambda(t)$, which carries the time dependence of the cosmological constant, appears linearly in the effective equations. Therefore in the first step of the computation it may be eliminated combining the two Friedmann equations, a system of two coupled equations for the pair $\{a(t),g(t)\}$ is obtained, and finally $\lambda(t)$ may be determined in the end of the computation.

\smallskip

It is also important to point out that scale-dependent gravity shares certain properties with other closely related approaches, such as the well-known running vacuum models \cite{RVM1,RVM2,RVM3,RVM4,RVM5,RVM6,RVM7}. Both methods take the 
idea of "running" seriously. However, scale-dependent gravity assumes that {\it{all}} couplings evolve with time, whereas in running vacuum models the main assumption is a running of the vacuum energy density only.

\smallskip

Another point to be mentioned here is the difference between scale-dependent gravity and Brans-Dicke theory \cite{BD}. Although in both theories an effective Newton's constant is present, they are completely different.
In scale-dependent gravity all quantities that enter into the action are promoted to scale-dependent quantities, which are allowed to vary with time, whereas in Brans-Dicke theory only Newton's constant varies via the evolution of the scalar field. 

\smallskip

Therefore, in the following we shall impose the following initial conditions
\begin{align}
&a(t_0)=1\,, \quad \dot a(t_0)=H_0\,,\label{idata1}\\
&g(t_0) = 1\,, \quad\dot g(t_0) = \dot{g}_0\,,\label{idata2} 
\end{align}
where $t_0$ is the age of the Universe, and in the following $\dot{g}_0$ is going to be the only free
parameter of the model. What is more, $\dot{g}_0 = 0$ corresponds to the classical solution (no running of Newton's constant), whereas $\dot{g}_0 \neq 0$ measures the deviation from the standard cosmological evolution.
Finally, the density parameters $\Wm$, $\Wr$, and $\Wl$ correspond to today's values, as in the usual $\Lambda$CDM case. 

The first Friedmann equation may be used as a constraint, which allows us to determine $\Omega_\Lambda$ if $\Omega_m$ and
$\Omega_r$ are specified. The normalized density corresponding to radiation is given by \cite{cardenas}
\begin{equation}
\Omega_r h^2 = 2.469 \times 10^{-5} (1+0.2271 \ N_{\text{eff}}),
\end{equation}
where $N_{\text{eff}}$ is the number of relativistic species, and $H_0 = 100 \:h~\textrm{km}/(\textrm{Mpc} \: \textrm{sec})$. In the numerical analysis we set $N_{\text{eff}}=3.04$ \cite{wmap}. Furthermore, following \cite{cosmo2} we set
\begin{equation}
H_0 = 73~\textrm{km}/(\textrm{Mpc} \: \textrm{sec}),
\end{equation}
and we vary the initial condition $\dot{g}_0$. 

\smallskip

\begin{figure*}[ht!]
\centering
\includegraphics[width=0.48\textwidth]{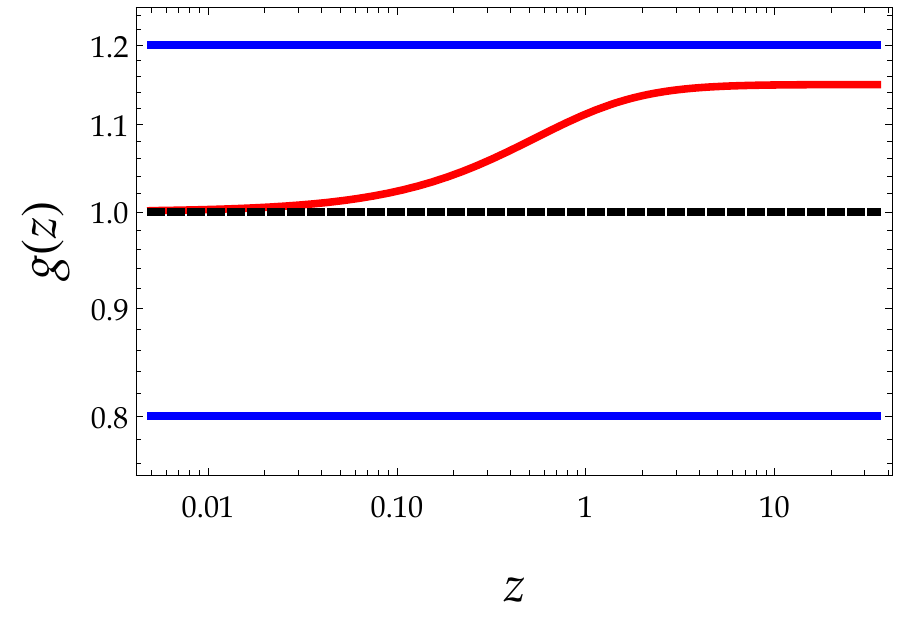}   \ 
\includegraphics[width=0.48\textwidth]{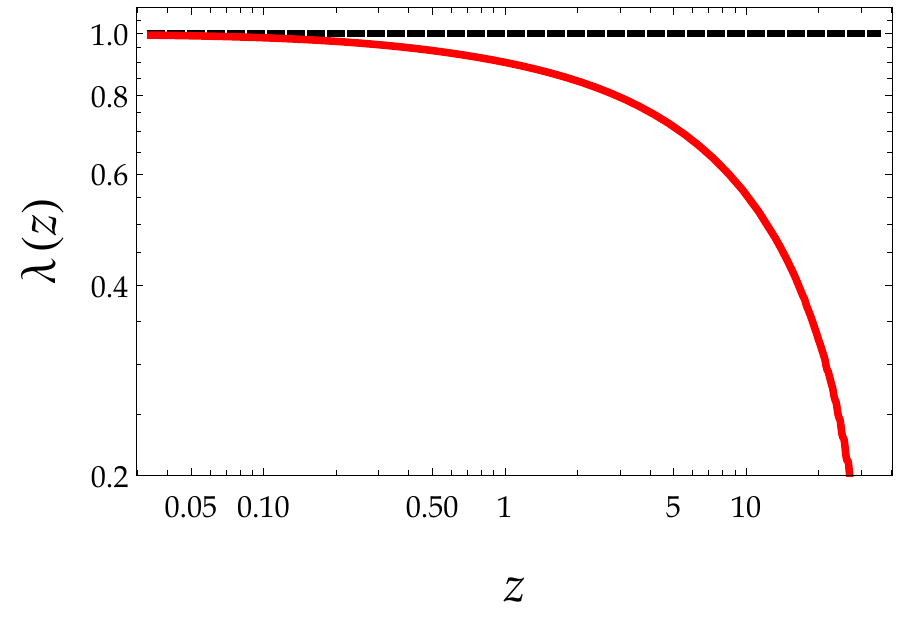}  \
\includegraphics[width=0.48\textwidth]{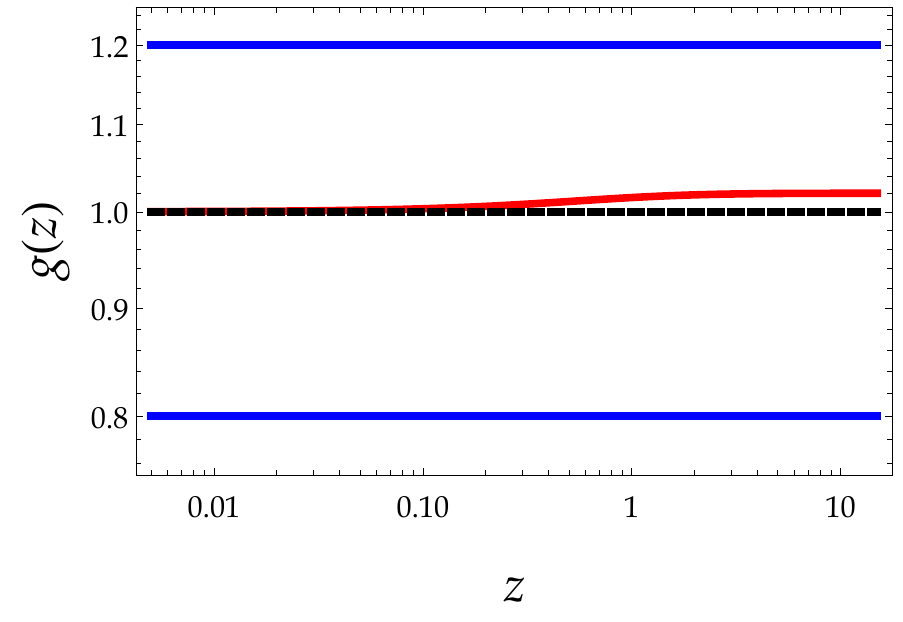}   \
\includegraphics[width=0.48\textwidth]{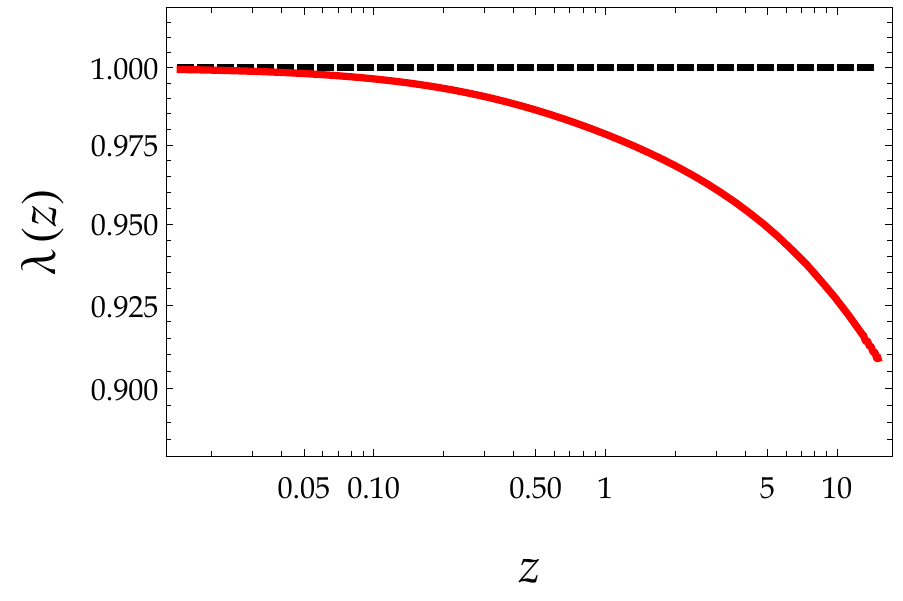}  \
\caption{ 
Dimensionless Newton's constant, $g(z)$ (first column), and dimensionless cosmological constant, $\lambda(z)$ (second column), versus $z$ for SN data set I (Gold, upper row) and for the SN data set II 
(Union2, lower row). In the case of data set I, $\Omega_m=0.309$ and $\dot{g}_0=-0.249 H_0$, whereas in the case of
data set II, $\Omega_m=0.269$ and $\dot{g}_0=-0.034 H_0$. Shown are: Numerical solution (red), classical values 
$g=1=\lambda$ (black) and the allowed strip from PBBN, $0.8 < g(z) < 1.2$ (blue).
}
\label{fig:2}
\end{figure*}

We should mention at this point that predictions of Primordial Big-Bang Nucleosynthesis (PBBN) for the abundances of the first light nuclei is sensitive to variations of Newton's constant. In this work we impose the constraint obtained 
in \cite{bbn}
\begin{equation}
\left|\frac{G_{\text{eff}}}{G} - 1\right| < 0.2.
\end{equation}
The behavior of the functions $g(t)$ and $\lambda(t)$ is shown in Fig. \eqref{fig:2}. In the early Universe $g(t)$ tends to a constant value different than unity, it is however compatible with the limits coming from PBBN.

\subsection{SN data}

The distance modulus $\mu=m-M$, where $m$ and $M$ are the apparent and absolute magnitude, respectively, is given by \cite{hogg,nesseris2004}
\begin{equation} 
\mu(z) = 25 + 5 \: \log_{10} \left[ \frac{D_L(z)}{\text{Mpc}} \right]
\end{equation}
where the luminosity distance, $D_L(z)$, is given by \cite{hogg,nesseris2004,brazil}
\begin{equation}
D_L(z) = (1+z) \int_0^z dx \frac{1}{H(x)}
\end{equation}
The free parameters of a given dark energy model are determined upon minimization of $\chi_{SN}^2$ computed by \cite{nesseris2004,negative,Ref1,Ref2}
\begin{equation}
\chi_{SN}^2 = X - \frac{Y^2}{Z}
\end{equation}
where the quantities $X,Y,Z$ are given by \cite{negative,Ref1,Ref2}
\begin{eqnarray}
X & = &  \sum_i^N \frac{(\mu(z_i)-\mu_i)^2}{\sigma_{\mu,i}^2}     \\
Y & = &  \sum_i^N \frac{\mu(z_i)-\mu_i}{\sigma_{\mu,i}^2}      \\
Z & = & \sum_i^N \frac{1}{\sigma_{\mu,i}^2}         
\end{eqnarray}
where all sums are over the number of points in the data set.
In this work we have used two data sets, data set I (Gold data set) with $N=157$ data points \cite{gold}, and data set II (Union2 compilation) with $N=557$ data points \cite{union2}.
When data set I is used, the $\chi^2$ minimization for the $\Lambda$CDM model yields $\Omega_m=0.309$, whereas when data set II is used, the $\chi^2$ minimization for the $\Lambda$CDM model yields $\Omega_m=0.269$. In the following we shall take those values as a prior, a strategy which is quite common in works on dark energy models, see e.g. \cite{nesseris2004,negative}.

In the case of the SD scenario, we find that $\chi^2$ minimization implies $\dot{g}_{0}=-0.249 H_0$ when data set I is used taking $\Omega_m=0.309$ as a prior. On the other hand, when data set II is used we find that $\chi^2$ minimization implies $\dot{g}_{0}=-0.037 H_0$ taking $\Omega_m=0.269$ as a prior. Therefore, in the discussion to follow, we shall use the following values to compute the evolution of matter perturbations
\begin{equation}
\Omega_m=0.309, \, \, \, \, \, \dot{g}_0=-0.249 \: H_0, \, \, \, \, \, \textrm{Gold data set}
\end{equation}
\begin{equation}
\Omega_m=0.269, \, \, \, \, \, \dot{g}_0=-0.037 \: H_0, \, \, \, \, \, \textrm{Union2 compilation}
\end{equation}

The value $\Omega_m$ is taken slightly lower than the Planck value, which poses no problem since its numerical value really depends on the model, see e.g. \cite{FengLi} for dark energy parameterizations where $\Omega_m$ was found to be significantly lower.

\smallskip

\begin{figure*}[ht!]
\centering
\includegraphics[width=0.48\textwidth]{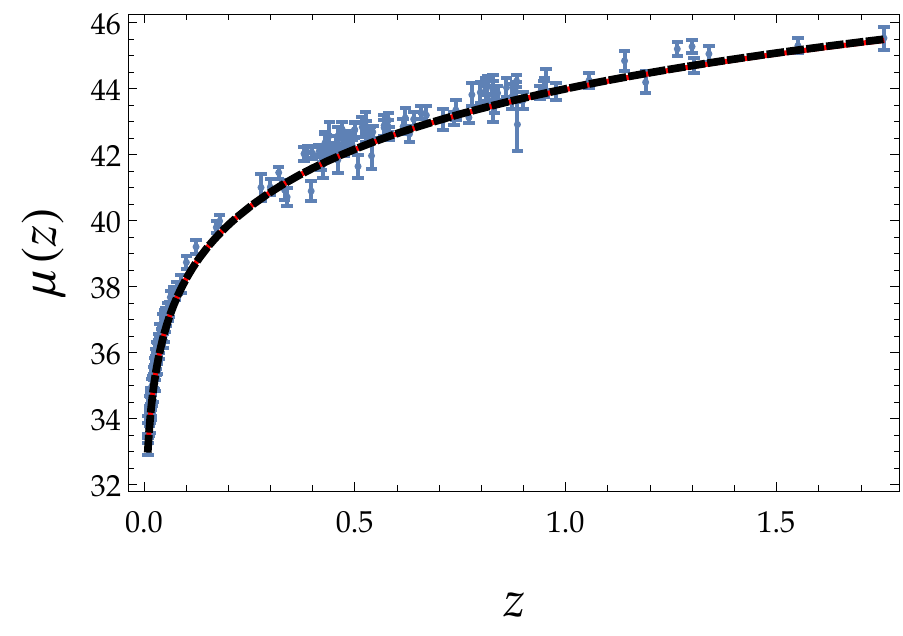}   \ 
\includegraphics[width=0.48\textwidth]{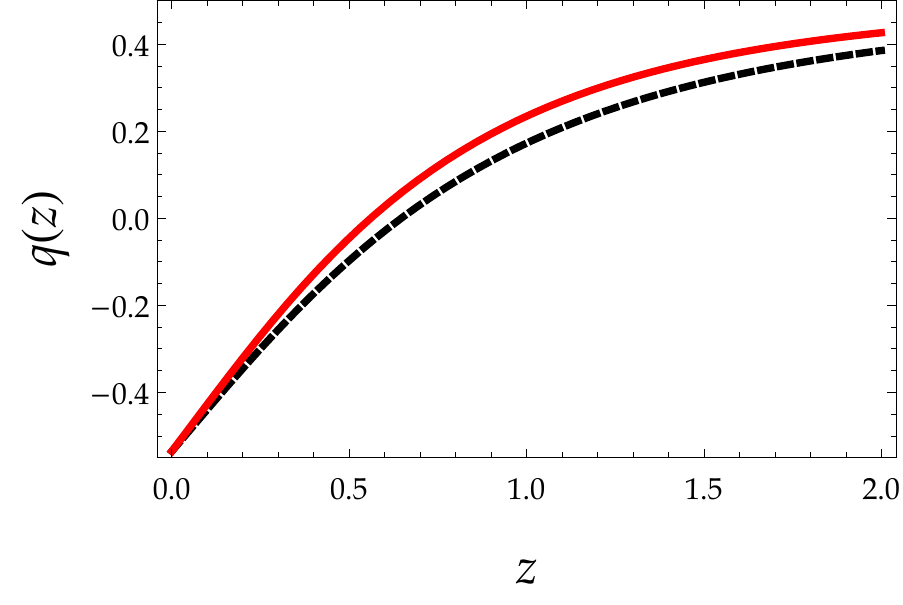}  \
\includegraphics[width=0.48\textwidth]{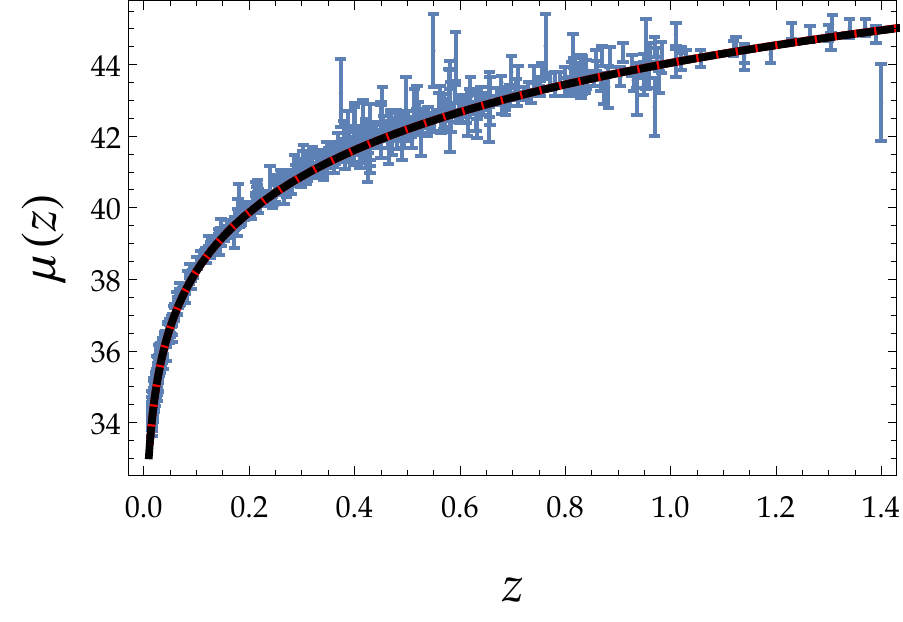}   \ 
\includegraphics[width=0.48\textwidth]{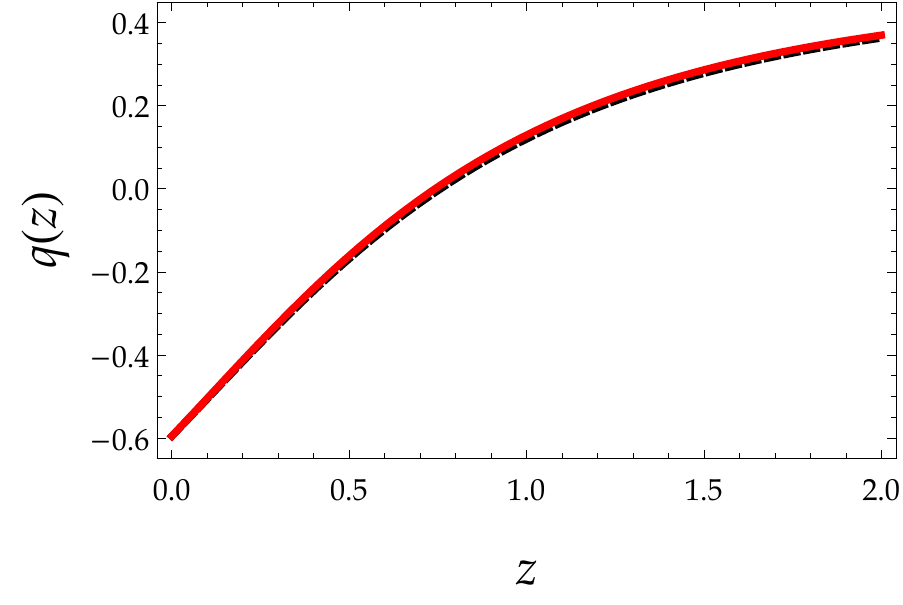}  \
\caption{ 
Distance modulus (first column) and deceleration parameter (second column) versus red-shift for SN data set I (Gold data set, upper row) and data set II (Union2 compilation, lower row). Black curves correspond to $\Lambda$CDM, and red curves to scale-dependent gravity.
}
\label{fig:1}
\end{figure*}

In Fig.~\eqref{fig:1} we show the distance modulus, $\mu(z)$, as well as the deceleration parameter, $q(z)$, which is 
given by \cite{nesseris2004,brazil}
\begin{equation}
q(z) \equiv - \frac{\ddot{a}}{aH^2} = -1 + (1+z) \frac{H'(z)}{H(z)}
\end{equation}
as a function of red-shift, $z$, for both SN data sets. The first row corresponds to data set I. Black curves 
correspond to $\Lambda$CDM, and red curves to scale-dependent gravity. In the second row, the impact of scale-dependent gravity is much weaker compared to the first case. In scale-dependent gravity the transition from deceleration to acceleration occurs at lower red-shift as compared to the $\Lambda$CDM model. In Table I we show the present value of the deceleration parameter, $q_0$, and the value of red-shift, $z_*$, at which the accelerating phase of the Universe starts
for scale-dependent gravity and for $\Lambda$CDM both for the Gold data set and the Union2 compilation.


\begin{table}[ht!]
\centering
\begin{tabular}{|c | c | c |} 
\hline
    Parameter & Gold & Union2 \\ [0.5ex] 
    \hline
   $q_0$  & -0.5363  & -0.5963  \\ 
    & (-0.5365) & (-0.5965)  \\
   \hline
   $z_*$ & 0.5613 & 0.7426 \\
    & (0.6475) & (0.7581) \\  
   \hline
\end{tabular}
\caption{
Deceleration parameter at present, $q_0$, and transition red-shift, $z_*$, for two cases: i) Scale-dependent gravity (without parenthesis), and ii) $\Lambda$CDM (with parenthesis) for the two data sets used in the present work.
}
\label{numerical_table}
\end{table}
  

\section{Matter density perturbations in scale-dependent gravity} 

Let us briefly review linear cosmological perturbation theory, see e.g. \cite{Mukhanov:2005sc,other}.

\smallskip

The goal is to solve the perturbed Einstein's field equations
\begin{equation}
\delta G_\nu^\mu = 8 \pi G \: \delta T_\nu^\mu
\end{equation}
On the one hand, for scalar perturbations relevant in growth of structures, the metric tensor has the form
\begin{equation}
ds^2 = -(1 + 2 \Psi) dt^2 + (1-2 \Psi) \delta_{ij} dx^i dx^j
\end{equation}
where $\Psi$ is the metric perturbation, while for the cosmological fluid the perturbed stress-energy tensor has the form
\begin{equation}
\delta T_0^0 = \delta \rho, \qquad \delta T_j^i = - \delta p \: \delta_j^i
\end{equation}
The full set of coupled equations for matter and metric perturbations may be found e.g. in \cite{mariam,pano1,pano2}.

\smallskip

The Fourier transform of the density contrast, $\delta_k = \delta \rho_m / \rho_m$, with $k$ being the wave number,
for pressure-less matter satisfies the following linear differential equation \cite{rogerio,leandros2,review2}
\begin{equation}
\ddot{\delta_k} + 2H \dot{\delta_k} - 4 \pi G \rho_m \delta_k = 0
\end{equation}
at linear level, $\delta_k \ll 1$, and for sub-horizon scales, $k/(2 \pi a) \gg a H$, when only non-relativistic matter clusters. During matter domination
\begin{equation}
a(t) \sim t^{2/3}, \; \; \; \; \; H(t) = \frac{2}{3 t}
\end{equation}
the matter density contrast  grows linearly with the scale factor, $\delta_k(a) \sim a$.

\smallskip

The equation for $\delta$ may be take equivalently the following form 
\begin{align}
\delta''(a) + \left( \frac{3}{a} + \frac{E'(a)}{E(a)} \right) \delta'(a) - \frac{3}{2} \frac{\Omega_m}{a^5 E(a)^2} \delta(a) = 0
\end{align}
where for simplicity we drop the sub-index $k$, a prime denotes differentiation with respect to the scale factor, 
while the dimensionless Hubble rate corresponding to $\Lambda$CDM can be written down as follows
\begin{align}
E(a)^2 \equiv \left(\frac{H(a)}{H_0}\right)^2 = \Omega_m a^{-3} + (1 - \Omega_m)
\end{align}
neglecting radiation at low red-shift, $1+z=a_0/a$, of order one or so.

\smallskip

In theories with a varying effective Newton's constant, $G_{\textrm{eff}}$, the equation for $\delta$ remains the same, the only difference being that $G$ is replaced by $G_{\textrm{eff}}$ \cite{L3,L5}. Thus, in scale-dependent gravity the
new equation to be solved is the following \cite{proDodelson,Dodelson,L4,L2}
\begin{align}
\delta''(a) + \left(\frac{3}{a} + \frac{E'(a)}{E(a)} \right)\delta'(a) - \frac{3}{2}\frac{\Omega_m}{a^5 E(a)^2} \left( \frac{G_{\text{eff}}}{G} \right) \delta(a) = 0
\end{align}
where $E = H / H_0$, as before, but now the Hubble parameter is the solution of the cosmological equation of scale-dependent gravity. Since the cosmological constant does not cluster, we need to solve a single fluid differential equation for the evolution of matter density contrast, and therefore $\Lambda(a)$ does not directly affects $\delta(a)$. Notice, however, that a varying cosmological constant affects the expansion of the Universe through the generalized Friedmann equations, and therefore it influences the evolution of $\delta(a)$ indirectly via $H(a)$.

\smallskip

\begin{figure*}[ht!]
\centering
\includegraphics[width=0.45\textwidth]{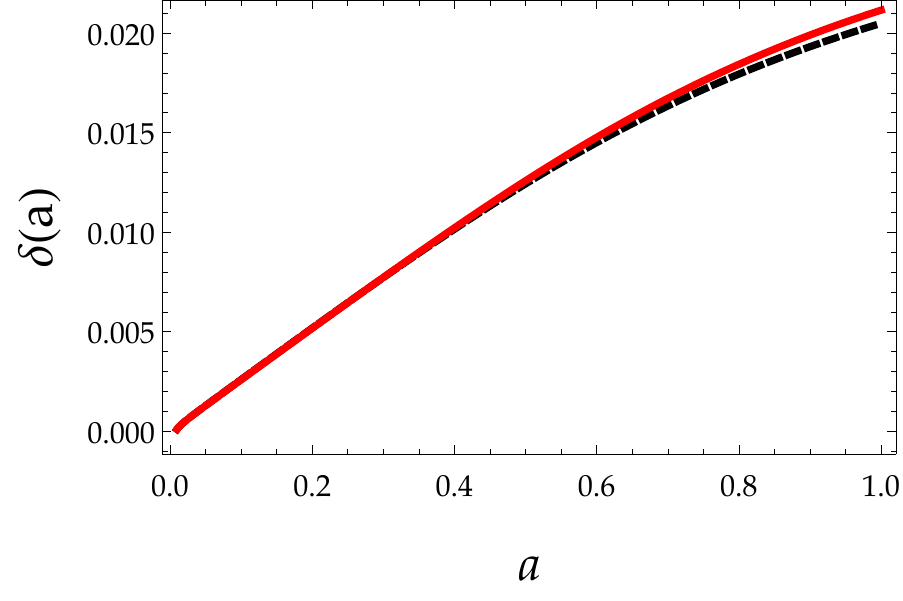}  \
\includegraphics[width=0.45\textwidth]{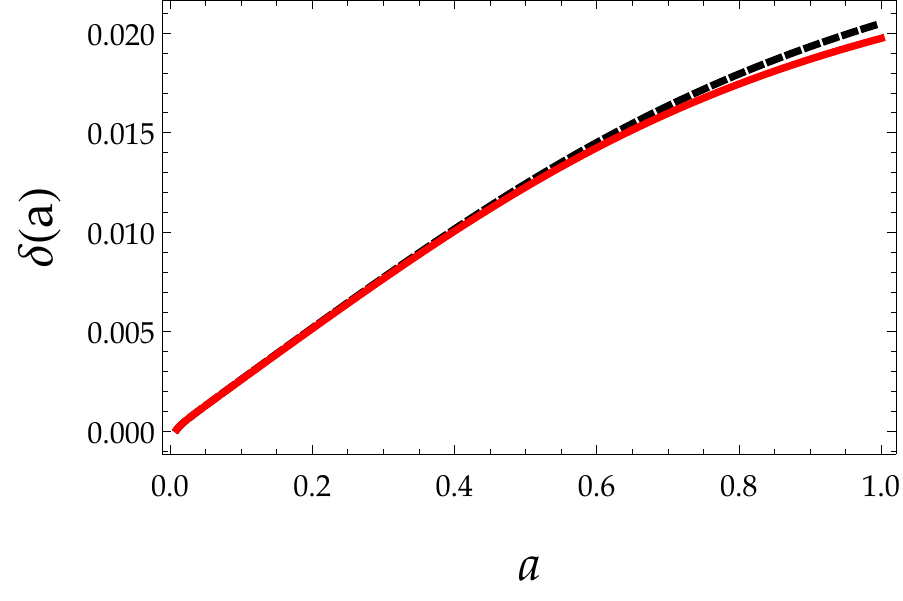} \
\caption{
Linear matter density contrast, $\delta(a)$, as a function of the scale factor, for the Gold data set (left panel) and the Union2 compilation (right panel). Black curves correspond to the $\Lambda$CDM model, and red curves correspond to scale-dependent gravity.
}
\label{fig:3}
\end{figure*}

In Fig.~\eqref{fig:3} we show the density contrast of non-relativistic matter versus the scale factor both for $\Lambda$CDM (black curves) and for scale-dependent gravity (red curves). The left panel corresponds to the Gold data set, while the right panel corresponds to the Union2 compilation. At early times, when matter dominates, all curves grow linearly with $a$ and they coincide, as expected. At late times, when dark energy starts to dominate, we observe that i) the $a$ dependence ceases to be linear, and ii) the curves split exhibiting different trends, i.e. in the case of Gold data set the density contrast for scale-dependent gravity lies above the one for $\Lambda$CDM, whereas for the Union2 compilation the opposite holds.

\smallskip

To make contact with observations, although there are several probes for dark energy, see e.g. the recent reviews \cite{review1,review2}, we shall consider here redshift-space distortions, and therefore we shall compute the combination \cite{L5,L1}
\begin{equation}
A(a) \equiv f(a) \sigma_8(a)
\end{equation}
where the first function, $f(a)$, is defined to be
\begin{equation}
f \equiv \frac{d \ln \delta}{d \ln a} = \frac{a}{\delta} \: \frac{d \delta}{d a}
\end{equation}
while the second function, $\sigma_8(a)$, is given by
\begin{equation}
\sigma_8(a) = \sigma_8 \frac{\delta(a)}{\delta(a=1)}
\end{equation}
with $\sigma_8 \equiv \sigma_8(a=1)$ being the rms density fluctuation on scales of $0.125 h^{-1} \: \text{Mpc}$.

\smallskip

The $\sigma_8$ values within scale-dependent gravity are obtained minimizing $\chi_{\text{SD}}^2$ corresponding to 
the $A(z)$ parameter
\begin{align}
\chi_{\text{SD}}^2 &= \sum _{i} ^{27}
\left(
\frac{ A_i - A_{\text{th}}(z_i)}{\sigma_{A,i}}
\right)^2
\end{align}
with $\sigma_A$ being the error of $A$ in the observational data taken from \cite{mariam}.

\smallskip

The value of $\sigma_8$ for $\Lambda$CDM extracted by the Planck Collaboration is found to be $\sigma_8 = 0.831$ \cite{planck1}, whereas upon comparison with the data used here we have obtained the following numerical values
\begin{equation} \label{arbitrary1}
\sigma_8 =
\left\{
\begin{array}{lcl}
0.72 & \mbox{ using }  & \mbox{ Gold data set}
\\
&
\\
0.77  & \mbox{ using }  & \mbox{ Union2 compilation}
\end{array}
\right.
\end{equation}
which are found to be lower than the Planck value.

\smallskip

\begin{figure*}[ht!]
\centering
\includegraphics[width=0.45\textwidth]{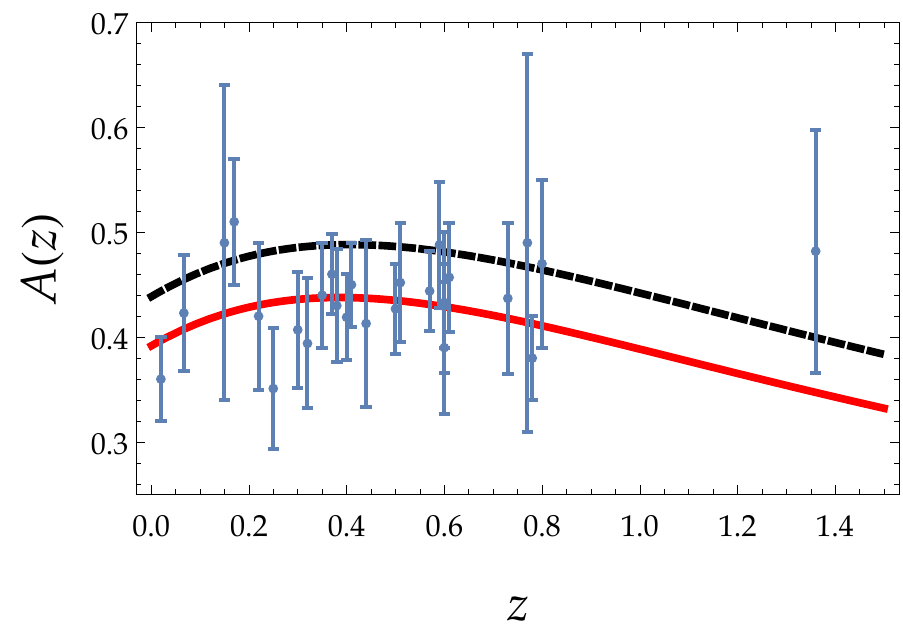}   \
\includegraphics[width=0.45\textwidth]{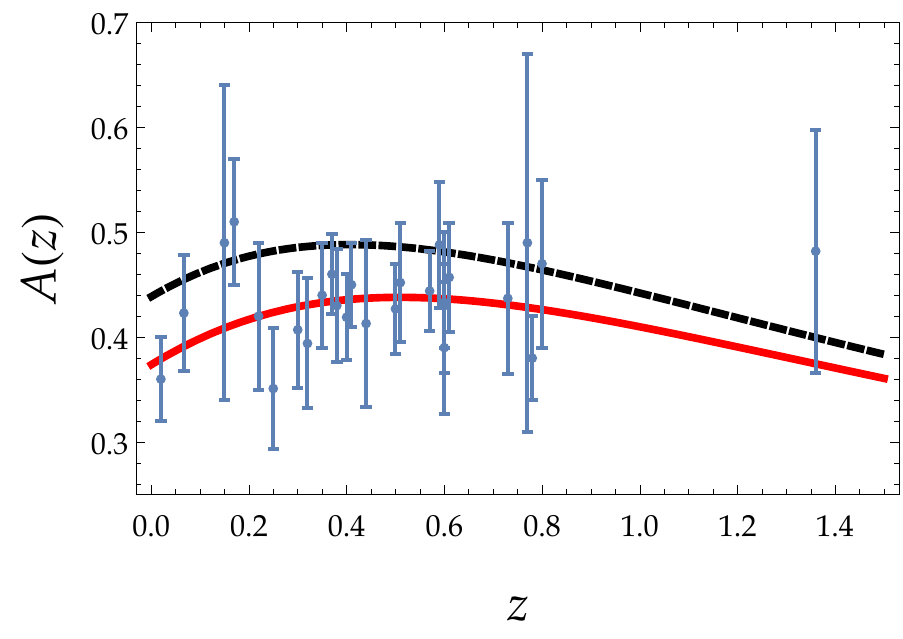}  \
\caption{ 
Combination $A=f \sigma_8$ versus $z$ for the Gold data set (left panel) and the Union2 compilation (right panel).
Black curves correspond to the $\Lambda$CDM model, and red curves correspond to scale-dependent gravity.
The observational points are shown as well.
}
\label{fig:4}
\end{figure*}

In Fig.~ \eqref{fig:4} we show the combination $A=f \sigma_8$ versus red-shift both for $\Lambda$CDM (black curves) and for scale-dependent gravity (red curves). We also show the observational points, taken from \cite{mariam}. 
As $\sigma_8$ was found to be lower in the case of the Gold data set, the deviation from the $\Lambda$CDM model is larger 
in this case.

%

Before we conclude our work, a couple of final comments are in order. Notice that we have not included more recent data $A(z)$, althought they do exist in the literature, since they are shifted towards higher red-shift, while it turns out that lower red-shift data are more adequate to probe $G$ variations corresponding to a weaker gravity at $z \sim 1$, as pointed out in \cite{L4}.

\smallskip

Finally, we have noticed that our findings are similar to the results reported in another closely related approach; the running vacuum scenario. In \cite{Gomez-Valent:2018nib}, for instance, the authors analyzing a running vacuum model 
computed the combination $A=f \sigma_8$, and they showed that in such a scenario the value of $f \sigma_8$ was slightly 
lower than its classical counterpart, in agreement with our results.


\section{Conclusions}

To summarize our work, in this article we have analyzed some implications of scale-dependent gravity to Cosmology. In particular, we have studied the evolution of the density contrast for non-relativistic matter within linear cosmological perturbation theory taking into account a varying Newton's constant. First, we solved the background cosmological equations using the first Friedmann equation as a constraint, which allowed us to impose the initial conditions properly and obtain self-consistent solutions. An effective gravitational coupling consistent with Primordial Big-Bang Nucleosynthesis was obtained. After that, we solved the generalized version of the equation for matter density contrast on top of a 
flat FLRW Universe, where Newton's constant was replaced by a varying effective gravitational coupling. Finally, with an eye on redshift-space distortion data and the $\Lambda$CDM crisis, the combination $A=f \sigma_8$ was computed. Next, it was contrasted against available observational data, and the numerical value of $\sigma_8$ that best fits the data was obtained. Our findings show that it is computed to be lower than that of $\Lambda$CDM extracted from the Planck Collaboration.

\section*{Acknowlegements}

We are grateful to the anonymous reviewer for a careful reading of the manuscript as well as for numerous useful comments and suggestions, which helped us significantly improve the quality of our work.
The author G.~P. thanks the Funda\c c\~ao para a Ci\^encia e Tecnologia (FCT), Portugal, for the financial support to the Center for Astrophysics and Gravitation-CENTRA, Instituto Superior T\'ecnico, Universidade de Lisboa, through the Project No.~UIDB/00099/2020 and grant No. PTDC/FIS-AST/28920/2017.
The author A.~R. acknowledges Universidad de Tarapac\'a for financial support.


\end{document}